\documentclass[9pt,twocolumn,twoside3]{osajnl}

\journal{ao} 

\setboolean{shortarticle}{false} 

\title{On the use of deep neural networks in optical communications}

\author[1]{Sanjaya Lohani}
\author[1]{Erin M. Knutson}
\author[2]{Matthew O'Donnell}
\author[3,*]{Sean D. Huver}
\author[1,+]{Ryan T. Glasser}

\affil[1]{Tulane University, New Orleans, LA 70118, USA}
\affil[2]{NG Next, Northrop Grumman Corporation, Redondo Beach, California 90278, USA}
\affil[3]{Deep Science AI, Eatontown, NJ 07724, USA}

\affil[*]{huvers@deepscience.ai}
\affil[+]{rglasser@tulane.edu}

\ociscodes{(100.4996) Pattern recognition, neural networks; (060.4510) Optical communications; (110.2970) Image detection systems.}

\begin{abstract}
Information transfer rates in optical communications may be dramatically increased by making use of spatially non-Gaussian states of light.   Here we demonstrate the ability of deep neural networks to classify numerically-generated, noisy Laguerre-Gauss modes of up to 100 quanta of orbital angular momentum with near-unity fidelity.  The scheme relies only on the intensity profile of the detected modes, allowing for considerable simplification of current measurement schemes required to sort the states containing increasing degrees of orbital angular momentum.  We also present results that show the strength of deep neural networks in the classification of experimental superpositions of Laguerre-Gauss modes when the networks are trained solely using simulated images.  It is  anticipated that these results will allow for an enhancement of current optical communications technologies.
\end{abstract}

\setboolean{displaycopyright}{true}

\begin{document}
\maketitle

\section{Introduction}
Optical communication relies on the generation, transmission, and detection of states of light to encode and send information.  While numerous protocols have been devised in order to increase the information transfer rate in optical communication scenarios, such as amplitude, phase, and quadrature-phase shift keying\cite{ASK,DPQPSK}, making use of the orbital angular momentum (OAM) degree-of-freedom of light is arguably one of the most promising methods\cite{OAMlg,OAM_1,Paterson1,Forbes2, Forbes3}.  For example, by generating, transmitting, and sorting states of light with OAM values of up to 14, bit transfer rates of $>\,1$ Terabit per second have recently been demonstrated\cite{Tbit,TbitFibers,Willner,freespace}.  As the number of quanta of OAM that an optical state may carry is fundamentally unlimited, current obstacles to even higher bit transfer rates are technical in nature \cite{Vienna}.  A primary technical difficulty is the accurate classification of OAM value detected at the receiving end of a communication platform\cite{Zeilinger300,ZeilingerSup,Zeilinger}. The conventional conjugate-mode sorting method requires a difficult optical alignment process and delivers consistently poor results for signals carrying more than a small amount of noise\cite{DosterWatnik}. Furthermore, the inaccuracy of this sort of method increases rapidly with increasing OAM quantum number, rendering high-OAM modes virtually impossible to classify.  Here we demonstrate the ability of deep neural networks to efficiently differentiate simultaneously between numerically-generated OAM states that have from 1 to 100 quanta of OAM with near-unity fidelity, even in the presence of substantial noise.  
Convolutional neural networks (CNNs) have recently been applied to the related task of demultiplexing combinatorially multiplexed OAM modes\cite{DosterWatnik}, with accuracies well exceeding those of the conjugate-mode sorting method\cite{freespace,Zeilinger}. While similar, this previous work differs from the present manuscript in the type of OAM-carrying beam (Bessel-Gauss versus Laguerre-Gauss), the nature of the images to train and classify (experimental only versus both experimental and simulated), and the network analysis. Furthermore, the method presented here relies only on the detection of the intensity profile of the OAM states, and bypasses technically-demanding protocols that are required to measure the phase profiles of the received modes\cite{Forbes4}.  We examine the effect of various network parameters on the classification accuracy, as well as differing sources of noise.  Lastly, we show that these networks can differentiate between numerous experimentally-generated superpositions of OAM modes with near perfect accuracy, when the networks are trained using only simulated images.
\subsection{Deep neural networks}
Deep neural networks (DNNs) have recently sparked a revolution in artificial intelligence due to their state-of-the art performance in fields such as computer vision, voice-to-text translation, and even gaming \cite{Neilsen,dnns}. 
Prior to 2012, deep neural networks were considered to be too computationally expensive and did not have the performance track record to be applied to practical scenarios.  This changed abruptly in 2012 when a DNN won the ImageNet computer vision classification competition \cite{ML1}.  This result, combined with the ability of graphical processing units (GPUs) to dramatically speed up neural network calculations via parallel processing, has resulted in significantly renewed interest in DNNs.  Additionally, the development of convolutional neural networks (CNNs) has allowed for a performance increase beyond previous neural network models that contain simply-connected neurons in successive layers \cite{ML2}.

\begin{figure}[H]
\centering\includegraphics[width=\linewidth]{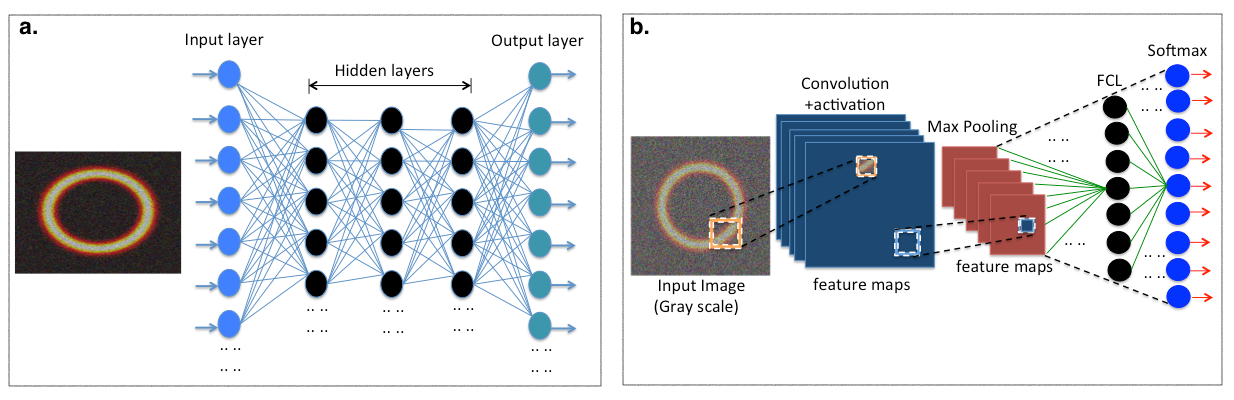}
\caption{Schematic of a \textbf{(a)} deep neural network and \textbf{(b)} convolutional neural network. A deep neural network consists of an input layer of neurons, multiple hidden layers, and an output layer, where each neuron is fully connected to the following layer and feeds the information forward to the output. A convolutional network consists of filter layers (including max pooling), and fully connected layers (FCL) as shown.}
\label{fig:neural_network}
\end{figure}
The framework of a general neural network consists of an input layer of neurons that each perform a nonlinear transformation on their respective inputs.  The output of each neuron is given a weight and bias, and the result is fed forward to the next layer of neurons.  This process is repeated until the final, output layer, which reaches a classification decision. 

In supervised learning, as used in the present manuscript, this output classification is compared to the known, desired result in order to calculate error. The error is minimized via a learning algorithm, and adjusted weights and biases are fed back to the neurons in the network.  One such learning iteration is termed an epoch.  After a chosen number of epochs, unknown test image(s) are then input into the network and classified at the output layer.  A deep neural network refers to artificial neural networks that contain more than one layer between the input and output layers (referred to as hidden layers). A schematic of such a network is shown in Fig. \ref{fig:neural_network}.

%
\subsection{Laguerre-Gauss states of light}
\begin{figure}[h!]
\centering\includegraphics[width=\linewidth]{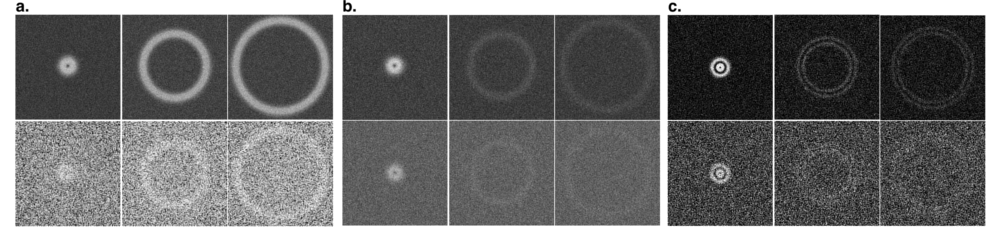}
\caption{Examples of numerically-generated, noisy non superposition LG images for the image sets \textbf{(a)} $S_{1}$, \textbf{(b)} $S_{2}$ and \textbf{(c)} $S_{3}$ which further have two subsets $S_{1A}$, $S_{2A}$, $S_{3A}$ (top row) and $S_{1B}$, $S_{2B}$, $S_{3B}$ (bottom row) each with $\ell$ values of 1 (left), 50 (center), and 100 (right). First two sets $S_1$ and $S_2$ are consist of non superposition OAM images at radial mode index p = 0, whereas set $S_3$ contains OAM images with radial mode index p = 1. Note that equal values of $\ell$ for the different data sets have differing radii, as the images in each data set are generated at different resolutions.  }
\label{fig:noisyLG}
\end{figure}

Laguerre-Gauss (LG) optical modes form the natural solutions of the paraxial wave equation in cylindrical coordinates.  These LG beams have an intensity null along their propagation axis, and have helical phase profiles that vary azimuthally as exp($i\ell\phi$), where $\ell$ is the quantum number associated with the degree of OAM that the state contains.  The integer $\ell$ corresponds to one $2\pi$ phase oscillation.  Importantly, the intensity spatial-distribution of LG modes with azimuthal index greater than, or equal to one consists of a single donut-like ring structure whose peak intensity radius scales as $\sqrt{\ell}$ (we note here that this is for radial mode index $p=0$).  Examples of numerically-generated, noisy LG modes for $\ell = 1,\,50,\,$and\,$100$ are shown in Fig. \ref{fig:noisyLG}.
There are a number of methods for measuring the degree of OAM ($\ell$) that an optical state carries.  As the phase fronts of LG modes contain $\ell$ quanta of $2\pi$ phase rotations, interferometry with a given LG state and a plane wave (or practically, a Gaussian beam) results in a fork pattern of interference fringes that scale with $\ell$\cite{ZeilingerSup}.  Alternatively, filters involving computer-generated holograms on spatial light modulators, in combination with single-mode fibers may be used to couple only to specific LG modes\cite{Zeilinger}.  All such methods make use of the phase-front structure of the LG modes that are to be detected.  Here we make use of the fact that the radius of the maximum intensity for a given LG mode scales as $\sqrt{\ell}$, and directly use (numerically-generated) intensity profiles as the inputs into our deep neural networks.  Additionally, the networks are trained using images that contain a varying amount of Gaussian noise, in order to simulate realistic experimental conditions, examples of which are shown in Fig. \ref{fig:noisyLG}.  In practice, it may be beneficial to use superpositions of LG modes to transmit information.  The intensity profile of superpositions of $\pm \ell$ LG modes consists of $2\ell$ bright (and dark) spots in a circular pattern, as discussed later.  We show that DNNs are powerful tools for differentiating between a variety of noisy, imperfect, experimentally-generated superpositions of LG modes.
\begin{table}[H]
\centering
\resizebox{\linewidth}{!}{%
\begin{tabular}{ |p{3cm}||p{4cm}|p{4cm}|p{4cm}|  }

 \hline 
 Network properties & Local 1 & Local 2 & VGG16\\
 \hline \hline
 Layers   &5&   varies & 16\\ \hline
 Convolutional  & yes   &no &yes\\ \hline
 Pre-trained  & varies&  no&yes\\ \hline
 Platform & Cypress supercomputer (or local CPU)& Cypress supercomputer &Deep Science AI GPU\\
 \hline
\end{tabular}}
 \caption{Networks described in this manuscript.}
 \label{tab: tab_1}
 \end{table}

\section{Methods}
\label{methods}
\subsection{Signal to noise ratio (SNR)}
The signal-to-noise ratio of the generated images is calculated as:
\begin{equation}
\textrm{SNR} = 10\times\textrm{log}_{10}(\mu_{s}/\sigma_{n}),
\end{equation}
where $\mu_{s}$ is the mean of the noiseless image pixel counts and $\sigma_{n}$ is the standard deviation of the added noise (i.e. the noisy image with the noiseless image subtracted). All SNRs are calculated with the images converted to grayscale, in order to quantify the intensity fluctuations. The SNR is calculated in this manner for consistency, as the different data sets were generated with separate resolutions and different amounts of added Gaussian noise. Note that the noise described in the following subsections \textbf{2. B} and \textbf{2. C} are additive noise.

\subsection{Numerically-generated non-superposition OAM modes}
In order to train and test the deep neural networks, we have generated non-superposition OAM images with $\ell\,=\,1$ to 100 and superposition modes between corresponding $\pm\, \ell$.  We have numerically generated 200 images for each LG mode (with radial mode index p = 0 and p = 1)  from $\ell=1$\,to\,100. The noiseless non-superposition LG images are generated using a modified version of the ``basic paraxial optics toolkit'' in Matlab\cite{toolkit}. 
We then add a variable amount of random Gaussian noise to each image, resulting in 200 images per value of $\ell$. We repeat this process to generate three total sets of images ($S_{1}$, $S_{2}$, and $S_{3}$) as shown in Fig. \ref{fig:noisyLG} to be used in the deep neural networks.
In the first set $S_1$, the peak intensity of each generated mode is normalized to a value of 255 (pixel value). The radial location of maximum intensity, and therefore the overall image SNR for a given amount of added noise, increases with increasing $\ell$ (since there is a larger region of noiseless intensity that is subtracted from the noisy images). 
The two generated series of images, $S_{1A}$ and $S_{1B}$, have mean intensity-noise values of 50.5 and 91.1 counts per pixel, respectively (again, the maximum intensity value a pixel may take is 255).  The standard deviation of the added noise for $S_{1A}$ is 28.2 counts per pixel, and 76.3 counts per pixel for $S_{1B}$.  The mean intensity and standard deviations are found by averaging over the values for all images with $\ell=$\,1, 25, 50, 75, and 100 in the respective sets of images. For the less-noisy data set $S_{1A}$, this corresponds to average SNRs from -3.81\,dB for $\ell=1$ to 2.77\,dB for $\ell=100$.  For the more noisy series of images $S_{1B}$, this corresponds to average SNRs that vary from -11.2\,dB for $\ell=1$ to -4.43\,dB for $\ell=100$.  Despite the variability of the SNR with $\ell$ due to the growing size of the LG modes, the visual image quality of the images stays quite constant, as seen in Fig. \ref{fig:noisyLG} \textbf{(a)}, where each image shown has a SNR equal to its $\ell$ value's average SNR (to within 0.02\,dB).

Next, the generated noiseless OAM images for each $\ell$\, = 1 to 100 in the set $S_2$ are normalized to same total intensity of $151,829$ pixel value (sum of all intensity pixel points). The two sets $S_{2A}$ and $S_{2B}$ have mean intensity-noise values of $35.33$ and $68.44$ counts per pixel, respectively. Similarly, the respective standard deviations of the added noise are $11.24$ and $15.64$ counts per pixel. Note that the noise added does not scale with the values of $\ell$. Hence, the less noisy set $S_{2A}$ and more noisy set $S_{2B}$ have normalized SNRs of $-2.12$\,dB, and $-3.57$\,dB, respectively. Lastly, the set $S_3$ contains OAM images with radial mode index $p\,=\,1$ for each $\ell\,=\,1$\, to 100. Again, the images are normalized to the same total intensity of $127,158$ pixel value. The mean intensity noise values and the standard deviations of the generated two sets $S_{3A}$ and $S_{3B}$ are $19.40$ and $56.03$ counts per pixel, respectively. Similarly, the respective normalized SNRs for the sets $S_{3A}$ and $S_{3B}$ are $-4.59$\,dB and $-8.70$\,dB. Here, the mean intensity and standard deviations of noise added, and the normalized SNR are found by averaging over the values for all images with $\ell=$\,1 to 100 in the respective sets of images. 
\subsection{Generating superposition OAM modes}
As with the non-superposition case, we have numerically simulated the noiseless superposition OAM images between $\pm\,\ell$ and then added randomly-distributed Gaussian noise to mimic the laboratory environment, in order to simultaneously classify 
the experimental OAM images. To be able to generate squeezed/elongated, elliptic, and twisted superposition OAM images at the beam waist (i.e. $z = 0$), we use the equations given by
\begin{eqnarray}\nonumber
|\psi_p^{\ell,-\ell}(r,\phi)|_{LG}^2 \propto r^{2|\ell|}\,L^{|\ell|}_p\Big(\frac{2r^2}{w^2}\Big)^2\,\exp\Big(\frac{-2r^2}{w^2}\Big)\,(1 \\
+\cos(2\,|\ell|\,\phi\,-\,\theta)) \,
\label{eqn:two}
\end{eqnarray}
and
\begin{eqnarray}\nonumber
|\psi^{n,-n}(r,\phi) |_{BG}^2 \propto J_{|n|}(\beta r)^2\exp\Big(\frac{-2r^2}{w^2}\Big)\,(1\,\\
+\,(-1)^{|n|}\,\cos(2\,|n|\,\phi\,-\,\theta)) 
\label{eqn:three}
\end{eqnarray}
for Laguerre-Gauss and Bessel-Gauss cases respectively. Here $|\psi_p^{\ell,-\ell}(r,\phi)|_{LG}^2$ and $|\psi^{n,-n}(r,\phi) |_{BG}^2$ are the intensities of the superposition of the field of $\pm\,\ell$ and $\pm\,n$, $r = \sqrt{x^2/a^2+y^2/b^2}$ is the radial distance from the center axis of the beam ($a\,\neq\,b$ provides the ellipticity), $p$ is the radial mode index, $\phi = \arctan(\frac{y}{x})$ is the phase of a helix, $w$ is the waist diameter, $L_p^{|\ell|}$ is the associated Laguerre-polynomial, $J_n$ represents the $n^{th}$-order Bessel function of the first kind, $\beta$ is the scale factor, and $\theta$ is the phase difference between the two superposed OAM modes. The proportionality sign in the expression is due to the fact that the constant factor while plotting the images is ignored because all the noiseless images, for each $\pm\,\ell$ and $\pm\,n$, are normalized to the fixed intensity pixel value. 

In order to make predictions and simultaneously classify the experimental image set $E_1$ as shown in Fig. \ref{fig:noisy_exp} (top two rows: \textbf{(a)}, \textbf{(b)}, \textbf{(c)}, \textbf{(d)}), we have numerically generated a training set $S_4$ with 180 randomly squeezed/elongated, and twisted superposition LG\,-\,OAM images with $720\,\times\,576$ resolution for each value of $\pm\,\ell\,=\,\pm\,1$ to $\pm\,10$, for a total of 1,800 images, by using equation \ref{eqn:two} with $-1\,\le\,(x,y)\,\le\,1$, $p = 20$, $a = 1$, $b = 1.27$, $\theta$ varying between 0 to $\pi$, and $w$ randomly varying between 0.95 to 1.30, 0.89 to 1.00, and 0.86 to 0.875 for $\pm\,\ell \le \pm \,3$, $\pm\,4\,\leq\, \pm\,\ell\,\le\,\pm\,5$, and $\pm\,6\,\le\,\pm\,\ell\,\le\,\pm\,10$, respectively, some examples of which are shown in Fig. \ref{fig:noisy_exp} (bottom two rows: \textbf{(e)}, \textbf{(f)}, \textbf{(g)}, \textbf{(h)}). The total intensity of the generated images for each $\pm\,\ell$ is normalized to $10,584,339$ pixel value. The mean intensity value and standard deviation of the added Gaussian noise are $34.73$ and $57.01$ counts per pixel, respectively. This gives a normalized SNR of $-3.08$\,dB. The mean and normalized values are found by averaging the corresponding values for images of each $\pm\,\ell$. Additionally, note that the images are generated in high resolution to match the resolution of the experimental images, but are both (simulated and experimental images) then converted to $224\,\times\,180$ pixels to increase the computational efficiency. 

Similarly, we have generated two separate training sets, $S_5$ and $S_6$, to make the OAM value predictions for the extremely noisy 23 experimental OAM images, set $E_2$, some of which are shown in Fig. \ref{fig:lg_bg_traning} (top row: \textbf{(a)}). First, we generate a set $S_5$ containing 192 randomly squeezed/elongated, and twisted superposition LG-OAM images with $720\,\times\,576$ resolution for each value of $\pm\,\ell\,=\,\pm\,1$ to $\pm\,40$, for a total of 7,680, with the settings $-1\,\le\,(x,y)\,\le\,1$, $13\le p \le 20$, $a = 1$, $b$ randomly varying from 1.00 to 1.35, $\theta$ varying between 0 to $\pi$, and $w$ randomly varying between 0.18 to 0.24, examples of which are shown in Fig. \ref{fig:lg_bg_traning} \textbf{(b)}. Next, the training set $S_6$ contains 180 randomly oriented, squeezed/elongated, and twisted superposition BG-OAM images with $300\,\times\,300$ resolution for each value of $\pm\,n\,=\,\pm\,1$ to $\pm\,20$, for a total of 3,600, which are generated by using equation \ref{eqn:three} with the settings (-0.012 to -0.008)$\,\le\,(x,y)\,\le\,$(0.008 to 0.013), $a$ randomly varying from 0.75 to 0.76, $b = 1$, $\beta$ randomly varying from 3,480 to 3,600, $\theta$ varying between 0 to $\pi$, and $w$ = 0.05, examples of which are shown in Fig. \ref{fig:lg_bg_traning} \textbf{(c)}. The two sets $S_5$ and $S_6$ are normalized to a total intensity of 6,579,471 and 4,317,461 total pixel values, respectively. The mean noise intensity value and standard deviation for the set $S_5$ are 38.10 and 68.12 counts per pixel, respectively, and the corresponding values for the set $S_6$ are 15.57 and 22.76 counts per pixel. The normalized SNR values for the generated sets $S_5$ and $S_6$ are then -6.40\,dB and 3.19\,dB. 
Finally, the images in the set $S_5$, $S_6$ and experimental set $E_2$ are converted to $300\,\times\,300$ resolution. In order to increase the computational efficiency and network accuracy, we scale all the images so as to have zero mean and unit variance before feeding to CNN and DNN\cite{scikit-learn}.

\subsection{Experimentally-generated modes}

\begin{figure}[h]
\centering\includegraphics[width=\linewidth]{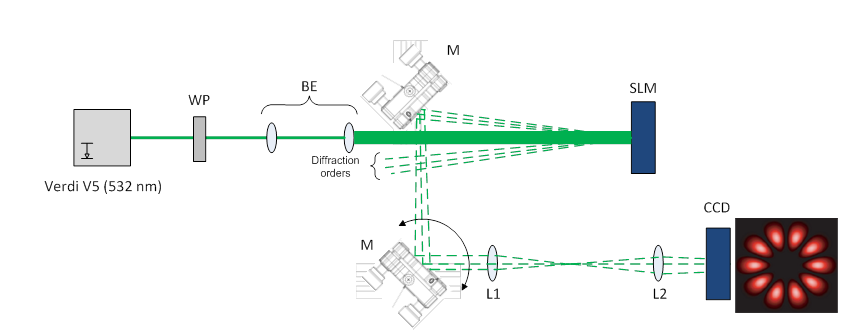}
\caption{Experimental setup to generate $\pm\,\ell$ superposition OAM modes.}
\label{fig:exp_setup}
\end{figure} 
A 532 nm laser beam produced by a Coherent Verdi V5 diode pumped laser is expanded (with the beam expander BE shown in Fig. 14) to illuminate a Holoeye PLUTO-NIR-011 Spatial Light Modulator (SLM) at normal incidence.  Diffracted vortex beams are picked off and directed through a set of lenses and into a CCD camera.  The SLM is a liquid crystal on silicon (LCOS) high-definition ($1,920\,\times\,1,080$ pixel with 8 $\mu$m pixel pitch) phase-only micro-display measuring $0.7$ inches diagonally. 
A phase between 0 and as much as $4.5\,\pi$ can be imparted to incident light by modulating the refractive index of liquid crystal voxels.  This is done by programming the SLM with a map of 8\,-\,bit voltages between 0 and 255 known as phase maps.  We generated phase maps corresponding to superpositions of two opposite-handed Laguerre-Gauss or Bessel-Gauss modes with orbital angular momenta of $\pm\,\ell$  as $1,920\,\times\, 1,080$ greyscale (8\,-\,bit) bitmap images. Illuminating these phase maps produces an interference pattern between the two vortex beams, forming a vortex structure in the far field known as petal pattern (bright petals) or a Ferris wheel (dark petals); some examples are shown in Fig. \ref{fig:noisy_exp}.  We multiplied each phase map by a blazed grating in order to diffract the vortex beam into the first diffraction order that we pick off from the illumination beam with the edge of a mirror as shown in Fig. \ref{fig:exp_setup}. A phase map multiplied by a blazed grating resembles the tines of a fork and is thus known as a ``forked grating." 

The diffracted intensity pattern is the Fourier transform of the product of the illumination beam's intensity profile and the SLM's reflectivity. Likewise, the phase profile of the diffracted intensity pattern is the Fourier transform of the product of the illumination beam’s phase profile and the SLM's phase profile, i.e. superimposed forked gratings. In our experimental setup lens L1 performed the Fourier transform, and lens L2 magnified the image onto a CCD camera. The lens L1 was placed at a distance from the SLM equivalent to its focal length of 300 mm. 

A spatial light modulator can be programmed with arbitrary phase maps to produce complex intensity patterns in the far-field, but not without distortion and artifacts. Because the SLM surface is pixelated, the actual blazed grating phase profile resembles a staircase which is equivalent to the original blazed grating convolved with a shallow high-frequency grating.  The Fourier transform of the actual blazed grating is a series of widely spaced “ghost” orders on top of orders produced by the blazed grating. In addition, not all illumination light is diffracted into the first order. Instead, some is lost to the zeroth order because the reflectivity of any liquid crystal interface varies with the liquid crystal refractive index. Sources of phase distortion can be attributed to wavefront error introduced by surfaces that make up the SLM and non-uniform distribution of liquid crystal depth across the active area of the SLM.

\subsection{Neural network activation function}
An activation function defines the logistic output at each node for a neural network's given input or sets of inputs. We use two different kinds of activation functions, sigmoid\cite{han1995influence} and Rectified Linear Unit (ReLU)\cite{glorot2011deep}.
The logistic sigmoid function ($\sigma$) maps every $\zeta \in [-\infty,+\infty]$ to [0,1] and is defined by 
\begin{equation}
\sigma(\zeta) = \frac{1}{1+\exp(-\zeta)},
\end{equation}
whereas the ReLU is a ramp function that takes only the positive part of the argument, such that 
\begin{equation}
 \text{ReLU}(\zeta) = \text{max}(0,\zeta). 
\end{equation} 
\section{Results}

 
 \begin{figure}[t!]
\centering\includegraphics[width=\linewidth]{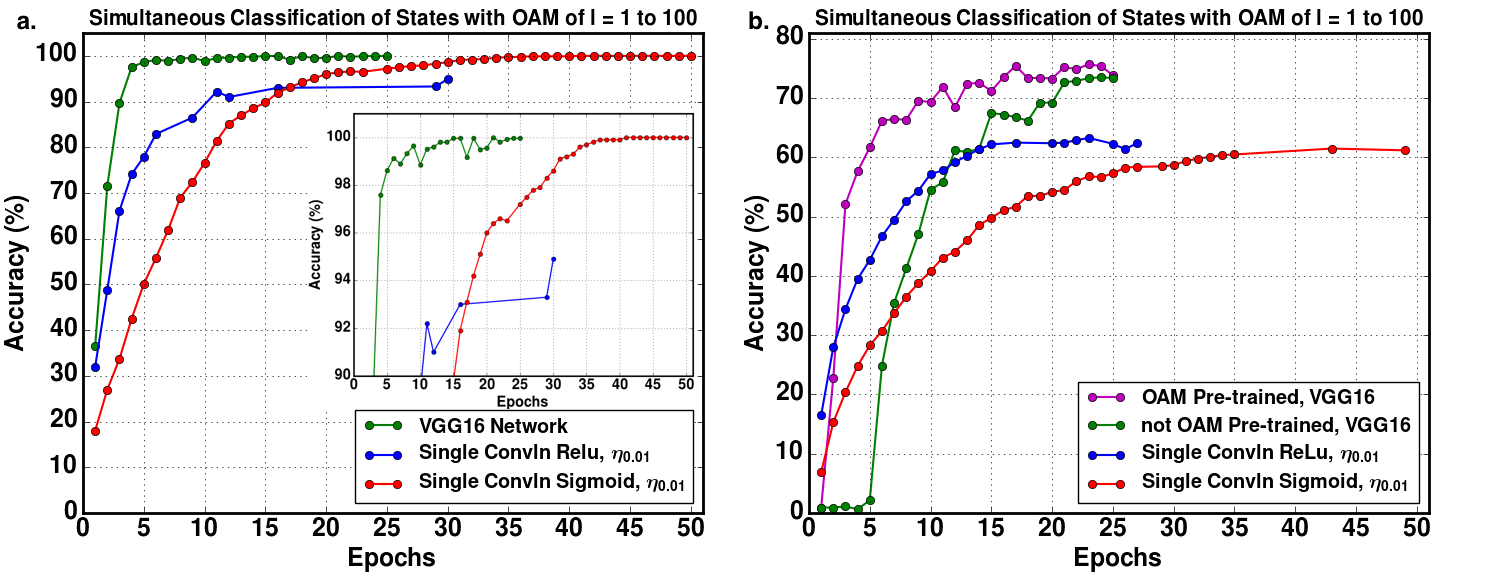}
\caption{OAM image classification accuracy of the locally-built CNN and deep convolutional VGG16 networks for the sets \textbf{(a)} $S_{1A}$ and \textbf{(b)} $S_{1B}$, described in detail in the text. For the VGG16 network, 2,525 randomly chosen images with $\ell$ values of 1 to 100 are used as test images, with the remaining 17,475 used as training images. For the locally-built CNN, we use a training set containing 18,000 images, and a validation set and test set with 1,000 images each. The network converges to near-unity classification accuracy for $S_{1A}$ after 5 and after 25 epochs for VGG16 and the local CNN, respectively. For the local CNN, we implement the ReLU (blue), and sigmoid (red) activation functions separately at different optimized learning rates $\eta$. The inset on the left shows the same data enlarged on an accuracy scale of $90\,-\,100\,\%$. For the noisier set, $S_{1B}$, the VGG16 network quickly rises to $>\,60\,\%$ accuracy after 5 epochs when the network is pre-trained with OAM images (magenta), whereas it learns at a slower rate when not pre-trained (green). 
}
\label{fig:s1_s2}
\end{figure}
Our chief results involve three separate networks as shown in Table \ref{tab: tab_1}.  First, we use a locally-built, 5-layer single convolutional neural network (CNN), which is run on Tulane University's Cypress supercomputer\cite{cyper}. A CNN is a special case of the general neural network described above, consisting of one or more convolutional layers (often with a pooling layer) which are followed by one or more fully connected layers as in a standard neural network\cite{cnn}. 
Here, we use a convolution of a 5\,$\times$\,5 filter with a single stride length (horizontal and vertical) followed by 2\,$\times$\,2 max pooling and fully connected layer. 
All the CNN classifications are based on three feature mappings per training image and cross-entropy error minimization\cite{entropy}. The training and test images used are not limited to a specific resolution. We vary the resolution according to the complexity of, and noise in, the OAM images. This flexibility allows for using experimental images of any given resolution. At the output, a fully connected layer (FCL as shown in Fig. \ref{fig:neural_network} \textbf{(b)}) with 200 fully connected neurons is attached to a ``softmax'' layer\cite{soft} which gives further probabilistic predictions. Being robust and of relatively small size, this network is computationally efficient, even at local CPU stations.

We then use a second custom-built network that is also run on Tulane University's Cypress supercomputer. This fully-connected deep neural network uses sigmoid neurons that allow for a small variation in the output sensitive to small changes in the weights and biases\cite{survey}.  The basic building blocks of this network -- the number of layers, number of neurons in each layer, and hyper-parameters -- are all designed to be externally and independently controllable. We are therefore able to manually vary individual parameters in order to quantify the network's classification accuracy dependence on each. We have optimized these parameters for OAM image classification accuracy, as a small change in any of these components may significantly affect the learning process. The number of input neurons is always equal to the size of the training and test images, and the output layer size is equal to the number of different OAM states that we are attempting to classify. The hidden layers process the data and transfer the information from one layer to the next, which is crucial for building a higher-level distributed network\cite{naf}.
\begin{figure}[b!]
\centering\includegraphics[width=\linewidth]{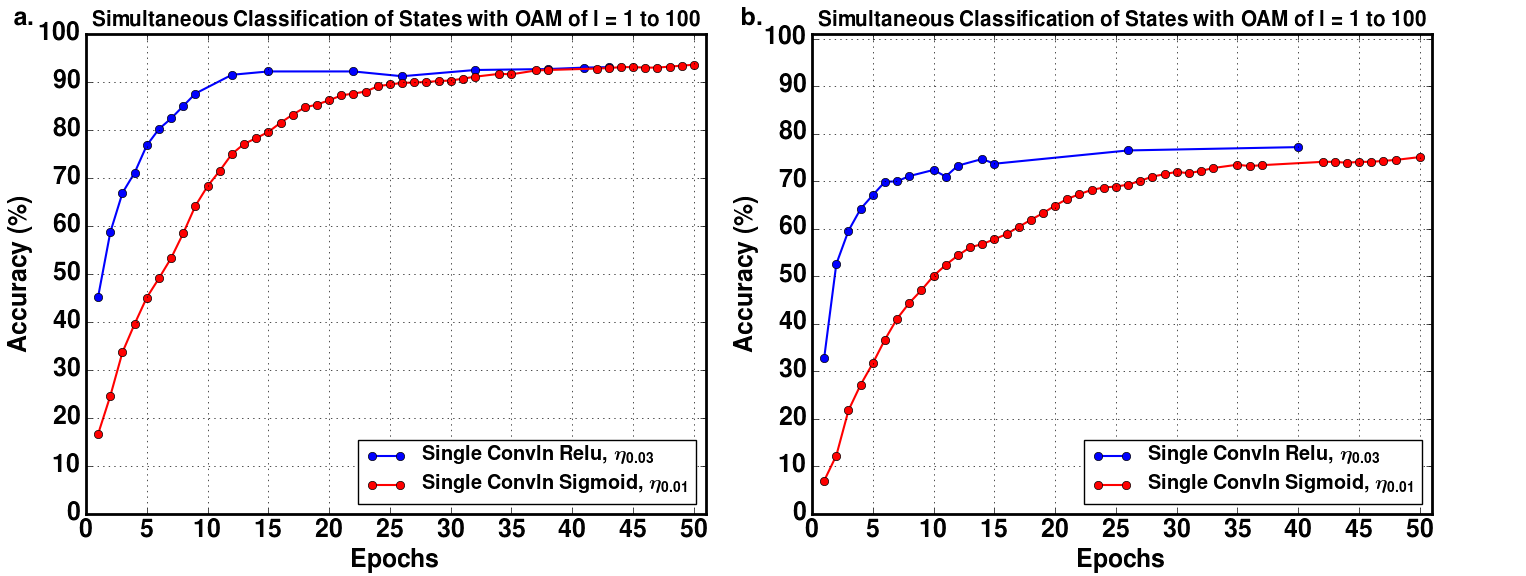}
\caption{Simultaneous classification accuracy of the CNN (``local 1'') for the SNR-normalized image sets \textbf{(a)} $S_{2A}$ and \textbf{(b)} $S_{2B}$. We again use 18,000 randomly selected images as the training set, and 1,000 images each for the validation and test sets.
}
\label{fig:2x}
\end{figure}

The third network we make use of is the VGG16 model, which is a 16-layer DNN that won first and second places for localization and classification tasks, respectively, in the 2014 ImageNet competition\cite{DCN}. We use a version of VGG16 that has been pre-trained on the ImageNet dataset, as it has been shown that pre-trained networks are able to learn features of new datasets quicker than those trained from scratch \cite{ML3}.
The training and test images used are fixed to a resolution of 224\,$\times$\,224 pixels.  This network is implemented on a Nvidia Titan X graphics processing unit at Deep Science AI.

The networks are trained stochastically until all the training sets are exhausted, after which we take the test sets and have the networks classify them correctly at the output. The accuracy is measured as the absolute percentage accuracy, the number of correctly predicted OAM images divided by the total number of test OAM images. This completes the first epoch, and the process continues until the last epoch. As expected, higher accuracies are generally reached at higher epochs. The result is a series of classification accuracies that are analyzed and plotted. Note that the training sets and testing sets used with the networks contain a uniform distribution of OAM images among the different OAM mode indices ($\ell$).
\begin{figure}[h!]
\centering\includegraphics[width=\linewidth]{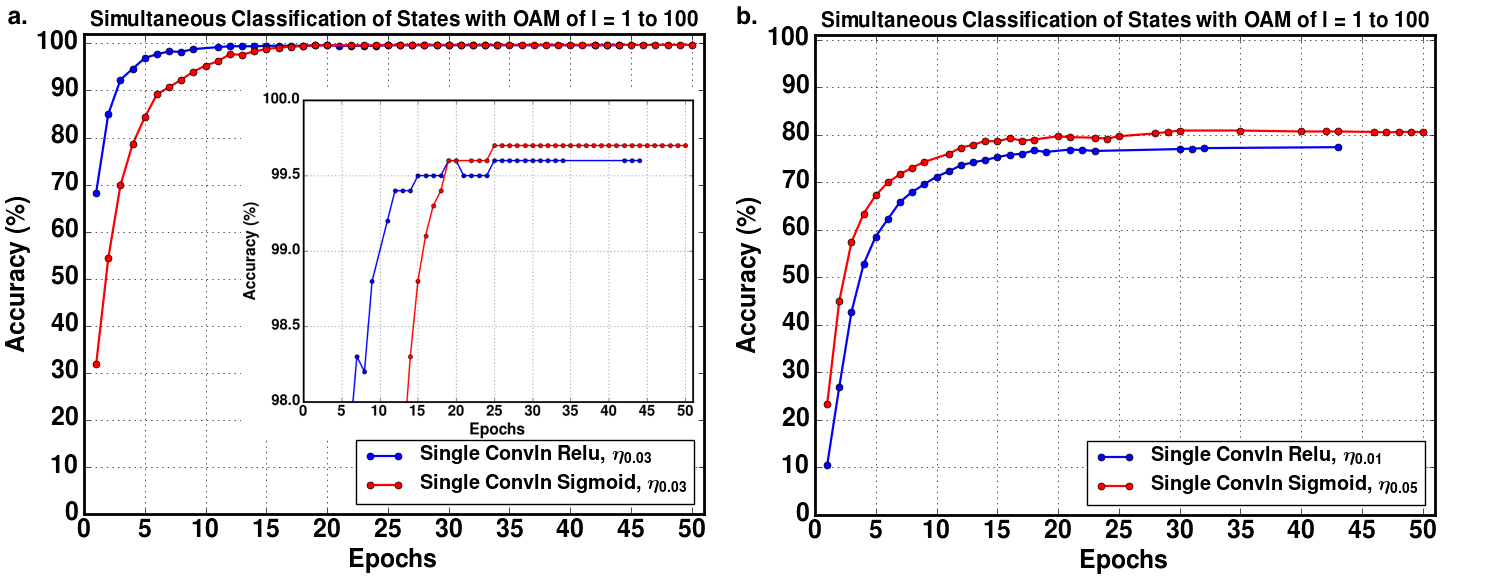}
\caption{Simultaneous classification accuracy of the CNN (``local 1'') for image set \textbf{(a)} $S_{3}$ with radial index number $p\,=\,1$. The network converges to near 99\,$\%$ classification accuracy for $S_{3A}$ after 7 epochs (ReLU) and 15 epochs (Sigmoid). The inset shows the same data enlarged on an accuracy scale of $98\,-\,100$\,$\%$. As expected, \textbf{(b)} the classification accuracy of the noisier set $S_{3B}$ is comparatively lower. The network reaches $>\,75\,\%$ accuracy after 10 epochs and achieves saturation at $\approx 81\,\%$ accuracy. }
\label{fig:p_1}
\end{figure}
\subsection{Simultaneous classification of OAM images}
We find that by using either the locally-built CNN, or the state-of-the-art VGG16\cite{DCN} network, near-unity classification accuracy of test images is rapidly achieved.  As seen in Fig. \ref{fig:s1_s2} \textbf{(a)},  the error 
of classifying 2,525 randomly chosen test images with OAM values of $\ell\,=\,1$\,to\,100 reaches 0.87\,$\%$ after only 6 epochs for the set of images $S_{1A}$ (the rest of the images from $S_{1A}$ are used for training). At epoch 21, the VGG16 network classified 100\,$\%$ of the randomly chosen test images correctly. The average error rate from epochs 5 to 25 is only 0.42\,$\%$, and the standard deviation of the error rate after 5 epochs is 0.43\,$\%$.  The computation time required per image classified is $\approx$\,10\,ms. Additionally, we have achieved 100\,$\%$ classification accuracy at epoch 41 with the locally-built CNN (using a lower resolution of $150\,\times\,150$ pixels and sigmoid activation at $\eta\,=\,0.01$), with an error rate of only 0.30\,$\%$ from epochs 30 to 41 . The local CNN makes predictions for the test images at the end of the epoch only when the classification accuracy for the validation set is greater than or equal to its previous value. This regularizes the output predictions of the network and increases the computational efficiency. This results in some gaps between predictions in the classification plot of the local CNN network.

Next we train and test the networks using the noisier set of images, $S_{1B}$.  As expected, the classification accuracy is lower than that achieved with the less-noisy image set.  Fig. \ref{fig:s1_s2} \textbf{(b)} shows that the VGG16 network reaches a classification accuracy of $>\,70\,\%$ for image set $S_{1B}$. We also find that the rate of increase in classification accuracy may be dramatically enhanced by starting from a network that has been pre-trained using the images in set $S_{1A}$. As seen in Fig. \ref{fig:s1_s2} \textbf{(b)}, this network pre-trained by the less noisy LG modes reaches a classification accuracy of $>\,60\,\%$ after only 5 epochs, whereas it takes 12 epochs to reach this level when not pre-trained on any LG OAM modes.  This is particularly promising with regards to the actual implementation of deep neural networks in practical optical communication schemes, as we see that pre-training with images of a different resolution (and noise) than might be used in a given experiment results in a significant improvement of the classification accuracy. Additionally, the local CNN network (with both ReLU and Sigmoid activation) saturates at  $\approx\,63\,\%$ accuracy.

\begin{figure}[h!]
\centering\includegraphics[width=\linewidth]{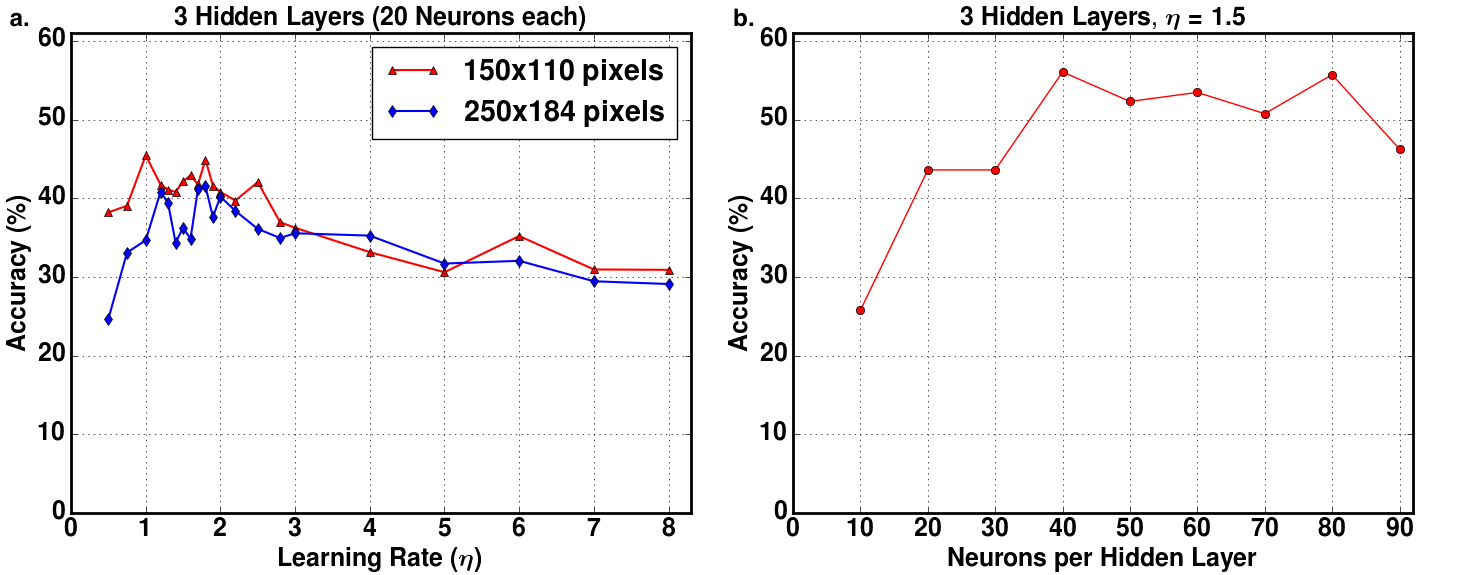}
\caption{\textbf{(a)} Maximum accuracy after 50 epochs versus learning rate for two training and test image sets of different resolutions, and \textbf{(b)} maximum accuracy versus number of neurons per hidden layer. These results are attained using the customized, non-convolutional deep net (``local 2'') with three hidden layers, and the image set $S_{1A}$.}
\label{fig:hidden_eta}
\end{figure}
Now we turn to using the intensity-normalized (see Methods), $150\,\times\,150$ pixel image set $S_{2A}$ with the locally-built CNN.  The simultaneous classification accuracy for images from set $S_{2A}$ with $\ell$\,=\,1 to 100 rises to $\approx\,80\,\%$ after 5 epochs and 20 epochs, and reaches saturation at $\approx \,94\,\%$ after 10 epochs and 30 epochs, when using the activation functions ReLU (blue) and Sigmoid (red), respectively, as shown in Fig. \ref{fig:2x} \textbf{(a)}. Similarly, for the noisier set $S_{2B}$, the CNN when using the ReLU activation function reaches $\approx 70\,\%$ accuracy after 5 epochs and converges to $77.2\,\%$ accuracy, whereas the sigmoid network reaches the same level of accuracy, albeit at a larger epoch, as shown in Fig. \ref{fig:2x} \textbf{(b)}.

\begin{figure}[b!]
\centering\includegraphics[width=\linewidth]{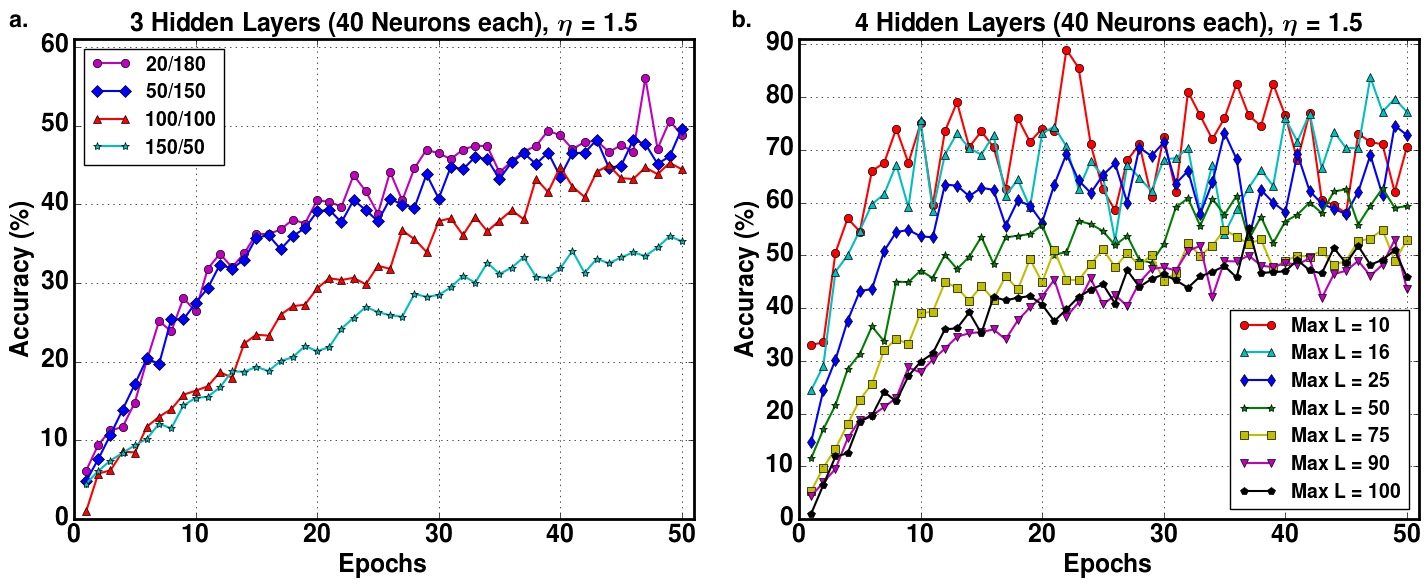}
\caption{Accuracy versus epochs for \textbf{(a)} varying ratios of testing images to training images used for each value of $\ell$, and \textbf{(b)} as the largest value of $\ell$ is varied, for the simultaneous classification of images.  Again, these results are for the non-convolutional, three-hidden-layer deep net (``local 2'') using image set $S_{1A}$.}
\label{fig:epochsmaxl}
\end{figure}
Finally, we test the CNN with image set $S_3$ which contain SNR-normalized LG-modes with radial index $p=1$ as shown in Fig. \ref{fig:noisyLG} \textbf{(c)}. This process uses the same number of training, validation, and test images as before. As shown in Fig. \ref{fig:p_1} \textbf{(a)}, the simultaneous classification accuracy of these OAM states with $\ell$\,=\,1 to 100 (all with $p=1$) reaches nearly $99\,\%$ accuracy after 7 epochs (ReLU) and 15 epochs (Sigmoid), and converges to $99.6\,\%$ and $99.7\,\%$, respectively, by epoch 25. The classification accuracy for the noisier set $S_{3B}$ is shown in Fig. \ref{fig:p_1} \textbf{(b)}. Here, the network performs similarly with either activation function, reaching $>75\,\%$ accuracy after 10 epochs and saturating at $\approx\,81\,\%$.

We now turn to the discussion of mapping out the parameter space to optimize deep neural networks for classifying noisy LG modes.  In order to accomplish this, we train our second network, the customized DNN ``local 2,'' with 180 images for each $\ell$ value of 1 to 100.  The network then simultaneously performs the classification of 20 images, again for each OAM value of 1 to 100 (i.e. 2,000 images in total).  This process is repeated as the parameters of the network are varied.

A crucial parameter in the performance of deep networks is the learning rate, $\eta$.  We have designed our network such that it finds the optimal learning rate for a given set of images (in all of the following, we use images from $S_{1A}$).  As such, shown in Fig. \ref{fig:hidden_eta} \textbf{(a)} we find that the optimal learning rate hyperparameter is approximately $\eta$\,=\,1.5.  We also note that the optimal learning rate is relatively robust to the resolution of the images that are used to both train and test the network, as Fig. \ref{fig:hidden_eta} \textbf{(a)} shows that the optimal learning rate stays close to $\eta\,=\,1.5$ for image resolutions of 250\,$\times$\,184 pixels, as well as for images that are 60\,$\%$ of this resolution.  Additionally, we find that the accuracy of the network tends to peak around 40 neurons in each hidden layer, as seen in Fig. \ref{fig:hidden_eta} \textbf{(b)}.

\begin{figure}[h!]
\centering\includegraphics[width=\linewidth]{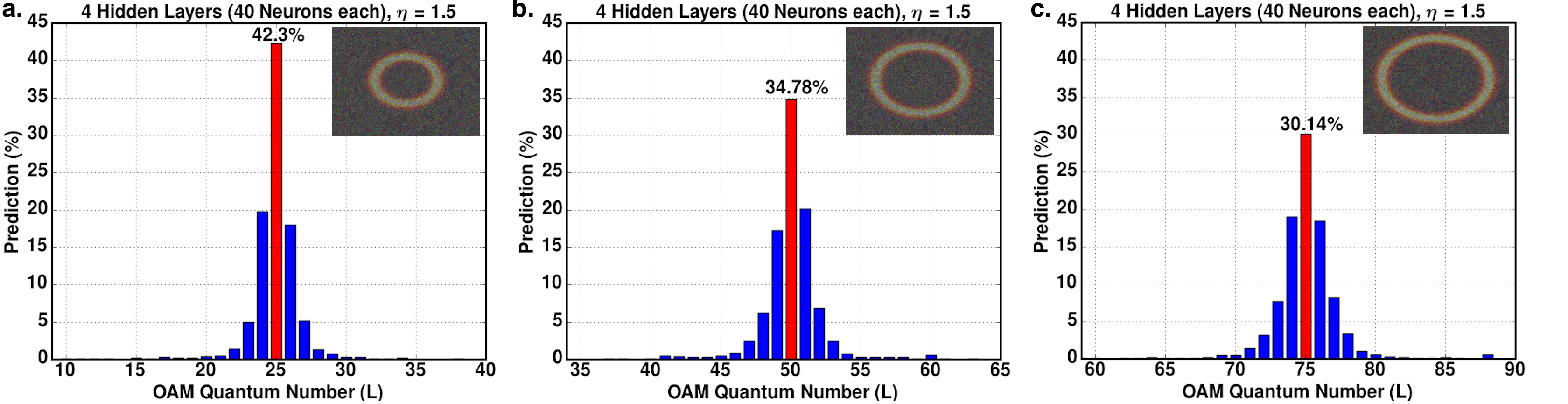}
\caption{Accuracy versus $\ell$ for individual testing of images, with $\ell$ values of \textbf{(a)} 25, \textbf{(b)} 50, and \textbf{(c)} 75. Again, these results are for the non-convolutional, four-hidden-layer deep net (``local 2'').}
\label{fig:single_L}
\end{figure}

By varying the ratio of testing images to training images used to train the network, we find that the optimal number of training images converges at approximately 150 images per value of $\ell$, as seen in Fig. \ref{fig:epochsmaxl}. We additionally investigate how the network's simultaneous classification efficiency varies as the maximum OAM quantum number, $\ell$, of the images to be classified increases.  As expected, a lower maximum value of $\ell$ results in a higher classification accuracy.
\begin{figure}[t!]
\centering\includegraphics[width=\linewidth]{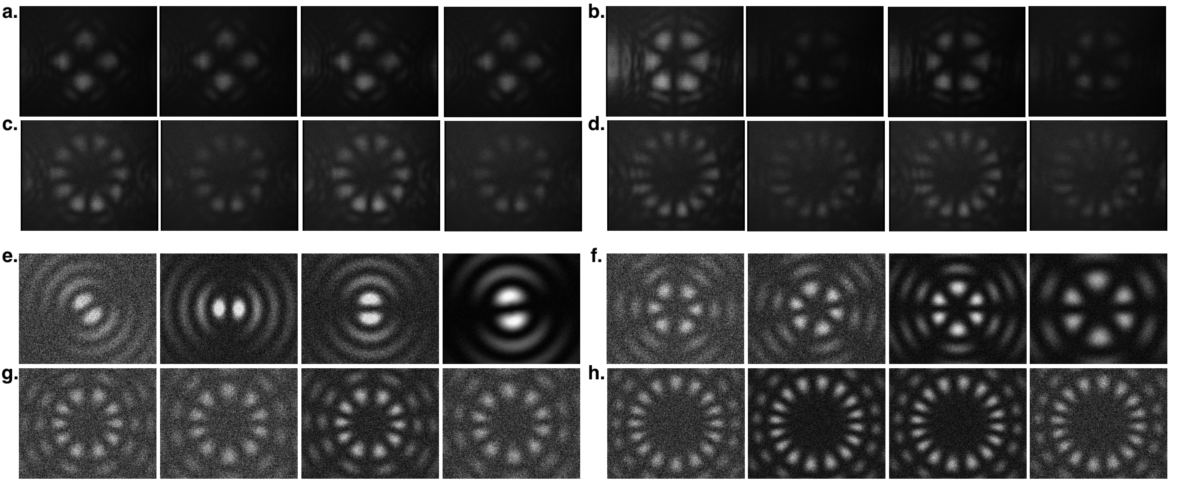}
\caption{Top: examples of experimentally-generated superposition OAM modes from set $E_1$ with \textbf{(a)} $\pm\,\ell\,=\,\pm\,2$, \textbf{(b)} $\pm\,3$, \textbf{(c)} $\pm\,5$, and \textbf{(d)} $\pm\,7$. Bottom: examples of numerically-generated, noisy superposition LG-OAM images from set $S_4$ with \textbf{(e)}$ \pm\,\ell\,=\,\pm \,1$, \textbf{(f)} $\pm\, 3$, \textbf{(g)} $\pm\,5$, and \textbf{(h)} $\pm\,8$ used in the training sets to classify $E_1$.}
\label{fig:noisy_exp}
\end{figure}

Lastly, in order to quantify how well the network classification scales with the specific LG mode order that is to be tested, we now classify individual test images, rather than large batches with many $\ell$ values simultaneously, after training the network as done previously. The same calculation is run for 100 trials, each with 50 epochs, in order to gain accuracy statistics. At the end of every epoch, the network predicts the value of $\ell$, resulting in 5,000 predictions for a given test image. Again as expected the accuracy of the network falls off with increasing $\ell$, though we importantly note that even for the largest LG modes used ($\ell$\,=\,100), the correct value is the most often classified by the network. Figure \ref{fig:single_L} shows typical results of the individual classification of LG images with $\ell$ values of 25, 50, and 75.

\subsection{Simultaneous classification of experimental OAM images}
\begin{figure}[b!]
\centering\includegraphics[width=.9\linewidth]{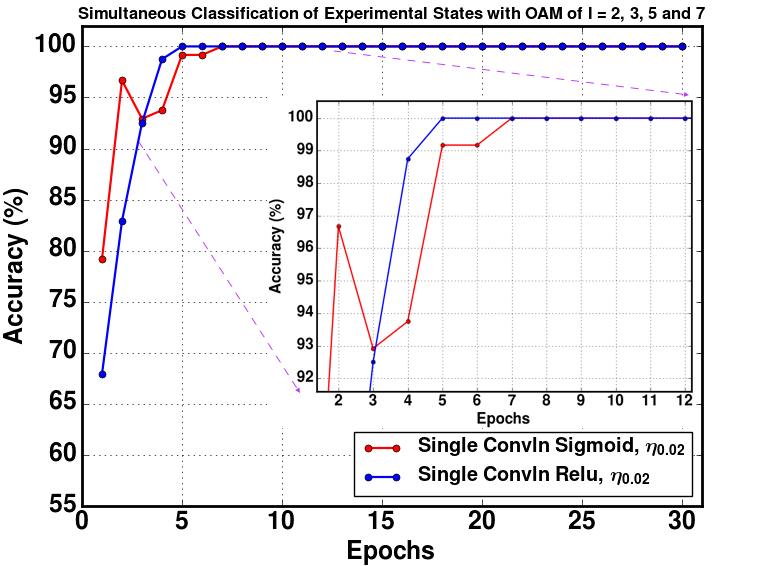}
\caption{Simultaneous classification of 240 experimental images of OAM modes with $\pm\,\ell\,=\,\pm \,2$, $\pm\, 3$, $\pm\, 5$ and $\pm\, 7$ (each with 60 different images) using a CNN network trained on simulated randomly oriented and squeezed/elongated images of OAM modes $\pm\,\ell\,=\,\pm\,1$\, to $\pm\,10$ (each with 180 different images). The experimental images have been simultaneously classified with unity fidelity in 5 epochs (blue) and 7 epochs (red) using  ReLU and Sigmoid activation functions at $\eta\,=\,0.02$, respectively. The same zoomed in classification results from epoch 2 to 12 are shown in the inset.}
\label{fig:matt_simul}
\end{figure}

We now turn to demonstrating the simultaneous classification of experimentally-generated superpositions of OAM modes (see the Methods section for a description of the experimental setup). First, we note that here the locally-built CNN is trained solely with simulated images.  These simulated superpositions consist of 180 numerically-generated randomly oriented and squeezed/elongated images for each value of $\pm\,\ell\,=\,\pm\,1$ to $\pm\,10$, for a total of 1,800 training images, each with $224\,\times\,180$ pixel resolution. Some examples are shown in the bottom of Fig. \ref{fig:noisy_exp}. The CNN then simultaneously classifies an experimental image set ($E_1$), examples of which are shown in the top of Fig. \ref{fig:noisy_exp}. This group of images is comprised of 4 sets of experimental images with OAM values $\pm\,\ell\,=\,\pm\, 2$, $\pm 3$, $\pm\, 5$, and $\pm\,7$ each with 60 different images (for a total of 240 OAM images).  The network very quickly reaches $100\,\%$ classification accuracy, as shown in Fig. \ref{fig:matt_simul}.  Note that in this case, there is no validation set or regularization at the output. The network always makes predictions for the test images at the end of each epoch. The CNN network with $\eta\,=\,0.02$ rapidly achieves 100\,\% accuracy in 5 and 7 epochs with the ReLU (blue line) and Sigmoid activation (red line) functions, respectively, as shown in Fig. \ref{fig:matt_simul}. The computation time required per image classified is $\approx\,8$\,ms.  We believe the limiting factor in computation time is the current networks' use of CPUs, and that making use of the large parallel-processing power of modern GPUs would result in a decrease in classification time.
\begin{figure}[h!]
\centering\includegraphics[width=\linewidth]{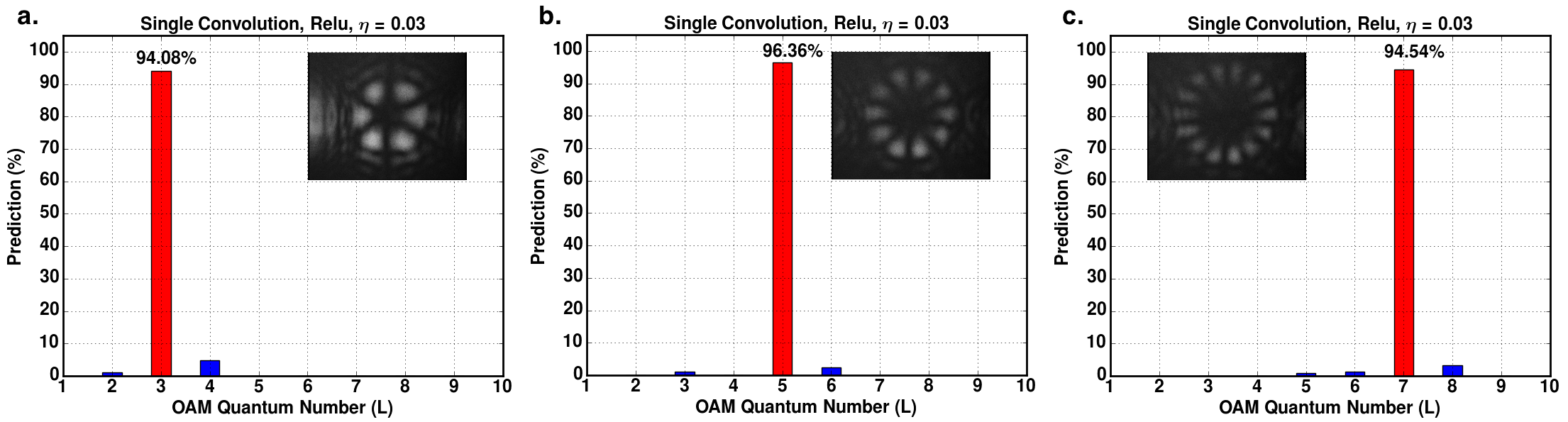}
\caption{Accuracy versus $\ell$ for individual testing of experimental images, with $\pm\,\ell$ values of \textbf{(a)} $\pm\, 3$, \textbf{(b)} $\pm\, 5$, and \textbf{(c)} $\pm\, 7$ with simulated images in the training sets of OAM value of $\pm\,\ell\,=\,\pm\,1$ to $\pm\,10$. The corresponding test images are shown.}
\label{fig:matt_single}
\end{figure}
Next we quantify the network's prediction accuracy of individual experimental OAM images, again trained only on the simulated images. The same calculation is run for 100 trials, each with 50 epochs, as described in subsection \textbf{3. A} with an experimental OAM as test image. 
The network has predicted the correct value of $\ell$ with an accuracy of $94.08\,\%$. $96.36\,\%$, and $94.54\,\%$  for a randomly-chosen experimental test image of OAM values $\pm\,\ell\,=\,\pm\, 3$, $\pm\, 5$, and $\pm \,7$, respectively, as shown in Fig. \ref{fig:matt_single} (inset). Using the adjusted Wald method\cite{wald,lewis}, we find accuracy intervals of 93.16\,$\%$\,-\,94.88\,$\%$, 95.61\,$\%$\,-\,96.99\,$\%$, and 93.65\,$\%$\,-\,95.31\,$\%$ for $\pm\,\ell\,=\,\pm\,3$, $\pm\,\ell\,=\,\pm\,5$, and $\pm\,\ell\,=\,\pm\,7$, respectively, with 99\,$\%$ confidence interval.

\begin{figure}[h!]
\centering\includegraphics[width=\linewidth]{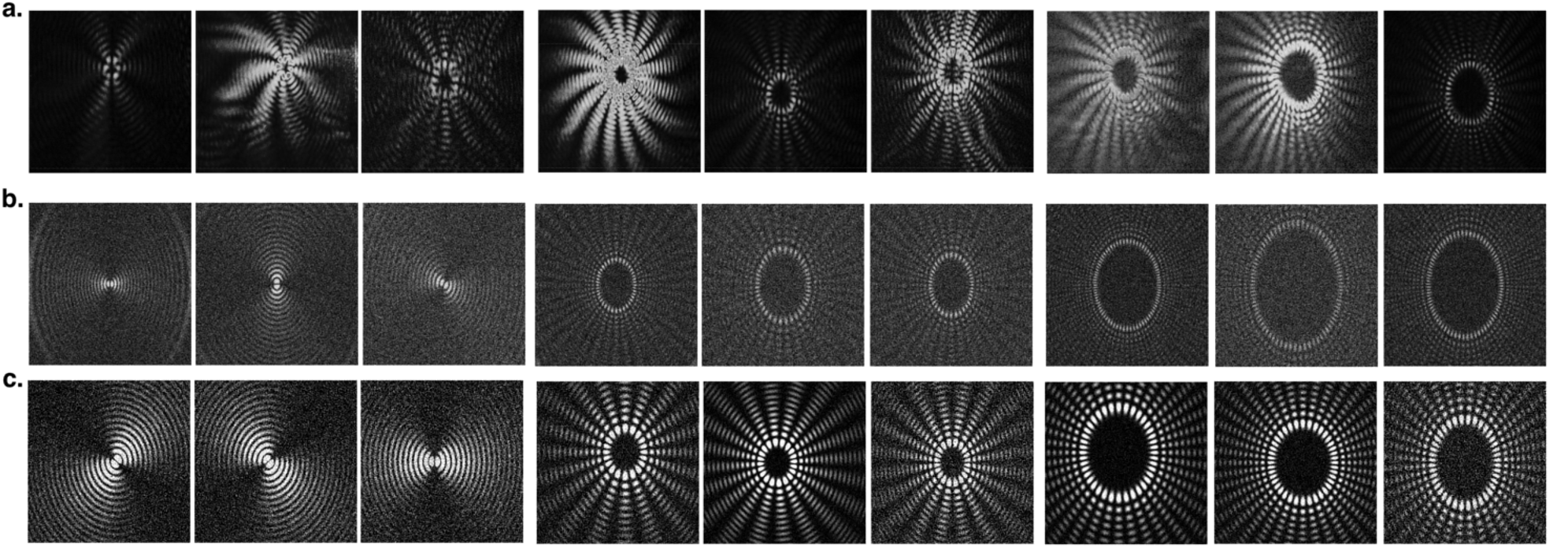}
\caption{Examples of the experimentally-generated extremely noisy OAM modes in the top row \textbf{(a)} (set $E_2$) with superpositions of $\pm\,\ell\,=\,\pm\,3$, $\pm\,4$, $\pm\,6$, $\pm\,8$, $\pm\,9$, $\pm\,10$, $\pm\,12$, $\pm\,16$, and $\pm\,18$ (from left to right, respectively), \textbf{(b)} examples of numerically-generated, noisy superposition LG images of $\pm\,\ell\,=\,\pm\, 1$, $\pm\,\ell\,=\,\pm\, 20$ and $\pm\,\ell\,=\,\pm\, 40$ (set $S_5$), and \textbf{(c)} numerically-generated BG images of $\pm\,n, =\,\pm\, 1$, $\pm\,n\,=\,\pm\, 10$ and $\pm\,n\,=\,\pm\, 20$ (set $S_6$), used in the training sets to make predictions for the given noisy experimental OAM set $E_2$.}
\label{fig:lg_bg_traning}
\end{figure}
\begin{figure}[h!]
\centering\includegraphics[width=\linewidth]{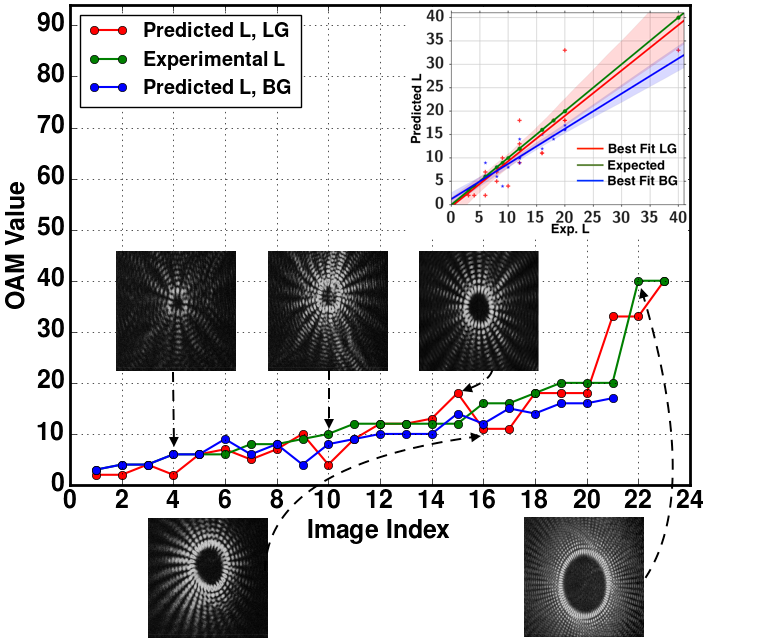}
\caption{Predicting the OAM value for the noisy experimental superposition images with the simulated LG (red line) and BG (blue line) images separately as the training set. Some examples of extremely noisy experimental images with comparatively larger prediction error are shown in the inset. Additionally, the regression fits, LG (red) and BG (blue), of the predicted results are also shown in the inset (top). 
}
\label{fig:lg_bg_prediction}
\end{figure}

Finally, we simultaneously classify an extremely noisy experimental image set $E_2$, examples of which are shown in Fig. \ref{fig:lg_bg_traning}, again using only simulated images in the training set. For this, we generate 192 and 180 noisy, randomly oriented, and squeezed/elongated LG and BG (Bessel-Gauss) images for each OAM value of $\pm\,\ell\,=\,\pm\,1$ to $\pm\,40$ and Bessel function order $\pm\,n\,=\,\pm\,1$ to $\pm\,20$, respectively, as shown in Fig. \ref{fig:lg_bg_traning}. We use these LG and BG sets separately as training images to make predictions for 23 experimental images. First, we train the networks with simulated LG modes from $\pm\,\ell\,=\,\pm\,1$ to $\pm\,40$ with large radial mode index $p\,>\,13$ (in order to mimic the effects of ringing on the experimental images) for 50 epochs and save the best configured network settings. Finally, we feed the experimental images to the network, resulting in the predicted OAM value as shown in Fig. \ref{fig:lg_bg_prediction} (red line) in $\approx 15$ ms. Note that this is distinct from previously explained individual test image prediction processes in which the network makes a prediction for the given single test image at the end of each epoch. Similarly, we perform the same steps with the simulated BG sets from $\pm\,n\,=\,\pm\,1$ to $\pm\,20$, with corresponding predictions shown in Fig. \ref{fig:lg_bg_prediction} (blue line). Again, note that we have used the Sigmoid activation function and hyper-parameter $\eta = 0.03$ to train the networks in the case of both the LG and BG training sets. In order to quantify the prediction accuracy of the network, we calculate the ``coefficient of determination,'' or $R^2$ score, which ranges from negative to 1. An $R^2$ of 1 indicates a perfectly predicted data set. We find $R^2$ scores of 0.82 and 0.77 for the network with LG and BG sets as the training images, respectively. Additionally, the figure in the inset shows the best fit lines for the predictions made by an LG (red) training set and a BG (blue) training set. The translucent bands around the fitted lines represent the $95\,\%$ confidence interval for the estimation of regression coefficients.

\subsection{Simultaneous classification of OAM images with multiplicative noise}
\begin{figure}[h!]
\centering\includegraphics[width=.9\linewidth]{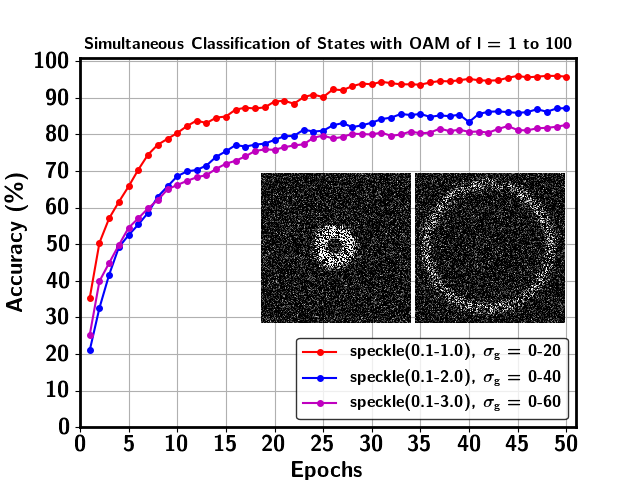}
\caption{Simultaneous classification accuracy of the CNN (``local 1'') for image set $S_{2A}$ (noiseless) with varying amount of speckle and Gaussian noise. The legend represents the amount of noise added to the image set, where for example, a speckle(0.1-1.0) represents the speckle noise with a variance ranging from square of 0.1-1.0 and $\sigma_g$ is the Gaussian noise. Examples of images with speckle(0.1-3.0), and $\sigma_g$ = 0-60 noise for OAM states of $\ell$ = 5 (left), and 100 (right) are shown in the inset.}    
\label{fig:speckle}
\end{figure}
We now turn to investigating the robustness of our CNN's ability to classify simulated images that contain multiplicative noise, in addition to Gaussian noise. A twofold approach is needed to mimic white noise as well as the noise that results from scattering and absorption effects. While Gaussian additive noise is appropriate for modeling unwanted signal modifications at the input and sensor, the noise arising from a communications channel is likely to be multiplicative\cite{multiplicative}.  Here we take noiseless 150\,$\times$\,150 pixels images from set $S_{2A}$ and add random multiplicative speckle and additive Gaussian noise, examples of which are shown in Fig. \ref{fig:speckle} (inset). We again use a local CNN network with $\eta$\,=\,0.005 and 18,000 images in the training set, and a validation set and test with 1,000 images each. We use three different combinations of speckle and Gaussian noise; speckle noise with variance ranging from square of 0.1 to 1.0 (less noisy), 0.1 to 2.0, and 0.1 to 3.0 (noisiest), with Gaussian noise ranging from $\sigma_g$\,=\, 0 to 20, 0 to 40, and 0 to 60, respectively. The simultaneous classification accuracy of OAM states of $\ell$ = 1 to 100 rapidly rises and saturates at $\approx$\,96\,$\%$ (red), 87.2\,$\%$ (blue), and 82.6\,$\%$ (magenta) as shown in Fig. \ref{fig:speckle}, respectively for the less noisy to noisiest image set.
\begin{table*}[h!]
\noindent\makebox[\linewidth]{\rule{\linewidth}{0.6pt}}
\centering
\resizebox{0.98\linewidth}{!}{
\begin{tabular}{| >{\centering\arraybackslash}p{6em}| >{\centering\arraybackslash}p{6em}| >{\centering\arraybackslash}p{6em} | >{\centering\arraybackslash}p{6em} | >{\centering\arraybackslash}p{6em} | >{\centering\arraybackslash}p{6em} |>{\centering\arraybackslash}p{6em} |>{\centering\arraybackslash}p{6em} |>{\centering\arraybackslash}p{6em} |>{\centering\arraybackslash}p{6em} |>{\centering\arraybackslash}p{6em} |}
\hline
Network& Training Set&Test Set&Resolution&OAM ($\ell$)& Classification\newline Mode&Radial Index&Type of\newline Noise&SNR\newline Normalization &SNR&Accuracy\\
\hline
\hline
CNN&simulated&simulated&150$\times$150&1 to 100&simultaneous&0&additive&no&-3.81 dB to\newline 2.77 dB&100\,$\%$\\
\hline
VGG16&simulated&simulated&224$\times$224&1 to 100&simultaneous&0&additive&no&-3.81 dB to\newline 2.77 dB&100\,$\%$\\
\hline
DNN&simulated&simulated&150$\times$110&1 to 100&simultaneous&0&additive&no&-3.81 dB to\newline 2.77 dB&55.10\,$\%$\\
\hline
CNN&simulated&simulated&150$\times$150&1 to 100&simultaneous&0&additive&no&-11.2 dB\newline to \newline -4.43 dB&63.3\,$\%$\\
\hline
VGG16&simulated&simulated&224$\times$224&1 to 100&simultaneous&0&additive&no&-11.2 dB\newline to \newline -4.43 dB&75.72\,$\%$\\
\hline
CNN&simulated&simulated&150$\times$150&1 to 100&simultaneous&0&additive&yes&-2.12 dB&93.6\,$\%$\\
\hline
CNN&simulated&simulated&150$\times$150&1 to 100&simultaneous&0&additive&yes&-3.57 dB&77.2\,$\%$\\
\hline
CNN&simulated&simulated&150$\times$150&1 to 100&simultaneous&1&additive&yes&-4.59 dB&99.7\,$\%$\\
\hline
CNN&simulated&simulated&150$\times$150&1 to 100&simultaneous&1&additive&yes&-8.70 dB&80.9\,$\%$\\
\hline
CNN&simulated&simulated&150$\times$150&1 to 100&simultaneous&0&additive + multiplicative&no&speckle (0.1-1) + $\sigma_g$ = 0 -20&96\,$\%$\\
\hline
CNN&simulated&simulated&150$\times$150&1 to 100&simultaneous&0&additive + multiplicative&no&speckle (0.1-2) + $\sigma_g$ = 0 -40&87.2\,$\%$\\
\hline
CNN&simulated&simulated&150$\times$150&1 to 100&simultaneous&0&additive + multiplicative&no&speckle (0.1-3) + $\sigma_g$ = 0 -60&82.6\,$\%$\\
\hline
CNN&simulated&experimental&224$\times$180&$\pm$ (2, 3, 5, 7) \newline (60 images each)&simultaneous&0\newline (superposition)&additive\newline(training set)&yes \newline (training set)&-3.08 dB\newline (training set)&100\,$\%$\\
\hline
CNN&simulated&experimental&224$\times$180&$\pm$3, $\pm$5, and $\pm$7&individual\newline prediction&0 \newline (superposition)&additive\newline(training set)&yes \newline (training set)&-3.08 dB\newline (training set)&94.08\,$\%$, 96.36\,$\%$, and 94.54\,$\%$, respectively\\
\hline
DNN&simulated&simulated&150$\times$110&25, 50, and 75&individual \newline prediction&0&additive&no&-3.81 dB to\newline 2.77 dB&42.3\,$\%$, 34.78\,$\%$, and 30.14\,$\%$, respectively\\
\hline
CNN&simulated\newline(LG and BG OAM)&experimental\newline(extremely noisy)&300\,$\times$\,300&various\newline(superposition)&simultaneous&$>13$\newline (training set)&additive\newline(training set)&yes\newline(training set)&-6.40 dB (LG)\newline 3.19 dB (BG)&$R^2$ score: \newline 0.82 (LG) and \newline 0.77 (BG)\\
\hline
\end{tabular}}
\caption{A brief summary of the classification of simulated and experimental LG-OAM by the networks.}
\noindent\makebox[\linewidth]{\rule{\linewidth}{0.6pt}}
\end{table*}

\section{Discussion}
We have demonstrated the ability of deep neural networks to simultaneously classify noisy, numerically-generated LG images containing OAM values of $\ell$\,=\,1 to 100 with error rates of less than 0.5\,$\%$ after 5 epochs.  Similar results of $>$\,99\,$\%$ accuracy are obtained by using states with nonzero radial index in order to increase the effective alphabet size.  Additionally, we find that we may increase classification ability substantially by pre-training the network with differing OAM image sets. We also demonstrate the ability of the networks to classify, with near-unity accuracy, experimentally-generated superpositions of OAM images.  By using a separate, customizable network, we have also investigated the dependence of the classification accuracy on various network parameters, including the learning rate, number of neurons per hidden layer, maximum $\ell$ value to be classified, and ratio of testing to training images used. Lastly, we have shown that deep convolutional neural networks may accurately and efficiently classify superpositions of OAM states of light, even under very noisy circumstances.  Results are summarized in Table 2.

We are optimistic that these results may be implemented in a realistic optical communication scheme by making use of recently developed ``squeezed nets''\cite{squeezenet}. Squeezed nets are DNNs that have been pruned and their weight matrices made sparse. They sacrifice a small amount of performance so their network description can be more compact ($<$\,10\,MB), and capable of being implemented on an FPGA (or ASIC), where on-board memory is typically scarce. We hope that the present demonstration will lead to the realistic implementation of increasing the information transfer rates of modern optical communications by allowing for substantial increases in usable alphabet size, with possible applications to quantum information schemes relying on superpositions of OAM states \cite{SPIE,TwistedLight,Forbes1,4photon,2photon}. 
\section*{Acknowledgements}
We acknowledge funding from the Louisiana State Board of Regents Research Competitiveness Subprogram under grant number 073A-15 and the National Science Foundation Graduate Research Fellowship under grant number DGE-1154145, as well as from Northrop Grumman -- NG NEXT.  We would also like to thank Jon D. Swaim and Onur Danac\i \hspace{0.5mm} for valuable discussions.  M.O'D would like to thank and recognize the valuable discussions regarding this work with colleagues David Yeaton-Massey and Stephane Larouche, both of NGNext.  

\bibliography{DNN_paper}

  \noindent

\end{document}